\documentclass[traditabstract]{aa}
\usepackage{graphicx}
\usepackage{txfonts}
\usepackage{natbib}
\bibpunct{(}{)}{;}{a}{}{,}

\newcommand{\Msun}{\ensuremath{\,{\rm M}_\odot}}           % Solar mass symbol
\newcommand{\Rsun}{\ensuremath{\,{\rm R}_\odot}}           % Solar radius symbol
\newcommand{\Teff}{\ensuremath{T_{\rm eff}}}               % Effective temperature symbol
\newcommand{\logg}{\ensuremath{\log g}}                    % log(g) symbol
\newcommand{\kms}{\,km\,s$^{-1}$}                          % km/s symbol
\newcommand{\Porb}{\ensuremath{P_{\rm orb}}}               % Stellar orbital period symbol
\newcommand{\EBV}{\ensuremath{E_{B-V}}}                                 % E(B-V) symbol
\newcommand{\mc}[1]{\multicolumn{2}{c}{#1}}
\newcommand{\reff}[1]{{#1}}                                  % makes corrections bold-face if wanted.

\begin{document} %%%%%%%%%%%%%%%%%%%%%%%%%%%%%%%%%%%%%%%%%%%%%%%%%%%%%%%%%%%%%%%%%%%%%%%%%%%%%%%%%%%%%%%%%%%%%
%%%%%%%%%%%%%%%%%%%%%%%%%%%%%%%%%%%%%%%%%%%%%%%%%%%%%%%%%%%%%%%%%%%%%%%%%%%%%%%%%%%%%%%%%%%%%%%%%%%%%%%%%%%%%%

\title{Orbital periods of cataclysmic variables identified by the SDSS.}

\subtitle{VI. The 4.5-hr period eclipsing system SDSS\,J100658.40$+$233724.4}

\author{John Southworth \inst{1} \and R.\ D.\ G.\ Hickman \inst{1} \and T.\ R.\ Marsh \inst{1}
        \and A.\ Rebassa-Mansergas \inst{1,2} \and B.\ T.\ G\"ansicke \inst{1}
        \and C.\ M.\ Copperwheat \inst{1} \and P.\ Rodr{\'i}guez-Gil \inst{3,4}
       }

\institute{Department of Physics, University of Warwick, Coventry, CV4 7AL, UK \ \ \ \ \ \email{jkt@astro.keele.ac.uk}
           \and Departamento de F\'{\i}sica y Astronom\'{\i}a, Universidad de Valpara\'{\i}so, Avenida Gran Bretana 1111,
                Valpara\'\i so, Chile
           \and Isaac Newton Group of Telescopes, Apdo.\ de Correos 321, E-38700, Santa Cruz de La Palma, Spain
           \and Instituto de Astrof\'{\i}sica de Canarias, V\'{\i}a L\'{\i}ctea, s/n, La Laguna, E-38205 Tenerife, Spain
          }

\date{Received ????; accepted ????}       % FORMAT: ???? is e.g. September 15, 1996 or March 16, 1997

\abstract{We present time-resolved spectroscopy and photometry of SDSS\,J100658.40$+$233724.4, which we have discovered to be an eclipsing cataclysmic variable with an orbital period of 0.18591324 days (267.71507 min). The observed velocity amplitude of the secondary star is $276 \pm 7$\kms, which an irradiation correction reduces to $258 \pm 12$\kms. Doppler tomography of \reff{emission lines from the infrared calcium triplet} supports this measurement. We have modelled the light curve using the {\sc lcurve} code and Markov Chain Monte Carlo simulations, finding a mass ratio of $0.51 \pm 0.08$. From the velocity amplitude and the light curve analysis we find the mass of the white dwarf to be $0.78 \pm 0.12$\Msun\ and the masses and radii of the secondary star to be $0.40 \pm 0.10$\Msun\ and $0.466 \pm 0.036$\Rsun, respectively. The secondary component is less dense than a normal main sequence star but its properties are in good agreement with the expected values for a CV of this orbital period. By modelling the spectral energy distribution of the system we find a distance of $676 \pm 40$\,pc and estimate a white dwarf effective temperature of $16500 \pm 2000$\,K. %SDSS\,J100658.40$+$233724.4 is an excellent target for high-speed photometry, which would allow precise measurements of the physical properties of this system.
}


%%%%%%%%%%%%%%%%%%%%%%%%%%%%%%%%%%%%%%%%%%%%%%%%%%%%%%%%%%%%%%%%%%%%%%%%%%%%%%%%%%%%%%%%%%%%%%%%%%%%%%%%%%%%%%
\maketitle %%%%%%%%%%%%%%%%%%%%%%%%%%%%%%%%%%%%%%%%%%%%%%%%%%%%%%%%%%%%%%%%%%%%%%%%%%%%%%%%%%%%%%%%%%%%%%%%%%%
%%%%%%%%%%%%%%%%%%%%%%%%%%%%%%%%%%%%%%%%%%%%%%%%%%%%%%%%%%%%%%%%%%%%%%%%%%%%%%%%%%%%%%%%%%%%%%%%%%%%%%%%%%%%%%

\section{Introduction}                                                                       \label{sec:intro}

Cataclysmic variables (CVs) are interacting binary systems containing a low-mass secondary star losing material to a white dwarf primary star. The Sloan Digital Sky Survey (SDSS) has spectroscopically identified 252 of these objects, 204 of which are new discoveries \citep[\reff{see}][and references therein]{Szkody+09aj}. This sample of SDSS CVs is valuable because of its large size and homogeneity \citep{Gansicke+09mn} and we are undertaking a project to characterise its constituent objects \citep[\reff{see}][and references therein]{Gansicke+06mn, Dillon+08mn, Me+06mn, Me+08mn, Me++08mn}. In the course of this work we have discovered that SDSS\,J100658.40$+$233724.4 (hereafter SDSS\,J1006) shows deep eclipses which are identifiable both spectroscopically and photometrically. The presence of eclipses allows us to determine the basic physical properties of the system, information which is difficult or impossible to obtain for the great majority of CVs \citep{SmithDhillon98mn,Knigge06mn,Littlefair+06sci}.

SDSS\,J1006 was discovered to be a CV by \citet{Szkody+07aj} on the basis of an SDSS spectrum which showed strong and wide Balmer emission lines. The continuum is blue at bluer wavelengths but clearly shows the spectral features of the M-type secondary star at redder wavelengths. SDSS\,J1006 is one of a select bunch of long-period CVs in which the eclipse of the white dwarf (WD) is directly visible in the light curve. In this work we present and analyse time-resolved spectroscopy and photometry of SDSS\,J1006, from which we measure the masses and radii of the WD and secondary star.

%%%%%%%%%%%%%%%%%%%%%%%%%%%%%%%%%%%%%%%%%%%%%%%%%%%%%%%%%%%%%%%%%%%%%%%%%%%%%%%%%%%%%%%%%%%%%%%%%%%%%%%%%%%%%%

\section{Observations and data reduction}                                           \label{sec:obs}

\begin{table*} \centering
\caption{\label{tab:obslog} Log of the observations presented in this work.
The mean magnitudes are calculated excluding observations taken during eclipse.}
\begin{tabular}{lccccccc} \hline
Date & Start time & End time & Telescope and & Optical  & Number of   & Exposure & Mean \\
\    &  (UT)      &  (UT)    &  instrument   & element  &observations & time (s) & magnitude   \\
\hline
2008 01 04 & 00:11 & 01:49 & WHT\,/\,ISIS        & R600B R316R gratings &  10 &    600 & \\
2008 01 04 & 04:43 & 05:50 & WHT\,/\,ISIS        & R600B R316R gratings &   7 &    600 & \\
2008 01 05 & 00:48 & 06:46 & WHT\,/\,ISIS        & R600B R316R gratings &  62 &    300 & \\
2008 02 01 & 02:31 & 05:41 & NOT\,/\,ALFOSC      & Wide-$V$ filter      & 197 & 10--60 & 18.8 \\
2008 03 14 & 00:06 & 03:07 & CAHA 2.2m\,/\,CAFOS & unfiltered           & 235 & 20--30 & 18.5 \\
2008 03 14 & 20:55 & 02:24 & CAHA 2.2m\,/\,CAFOS & unfiltered           & 199 & 20--30 & 19.0 \\
2008 03 15 & 22:56 & 02:18 & CAHA 2.2m\,/\,CAFOS & unfiltered           & 312 & 20--30 & 18.4 \\
2008 12 21 & 03:17 & 05:17 & NOT\,/\,ALFOSC      & Wide-$V$ filter      & 362 &     10 & 18.7 \\
\hline \end{tabular} \end{table*}

\begin{figure*} \centering
\includegraphics[width=\textwidth,angle=0]{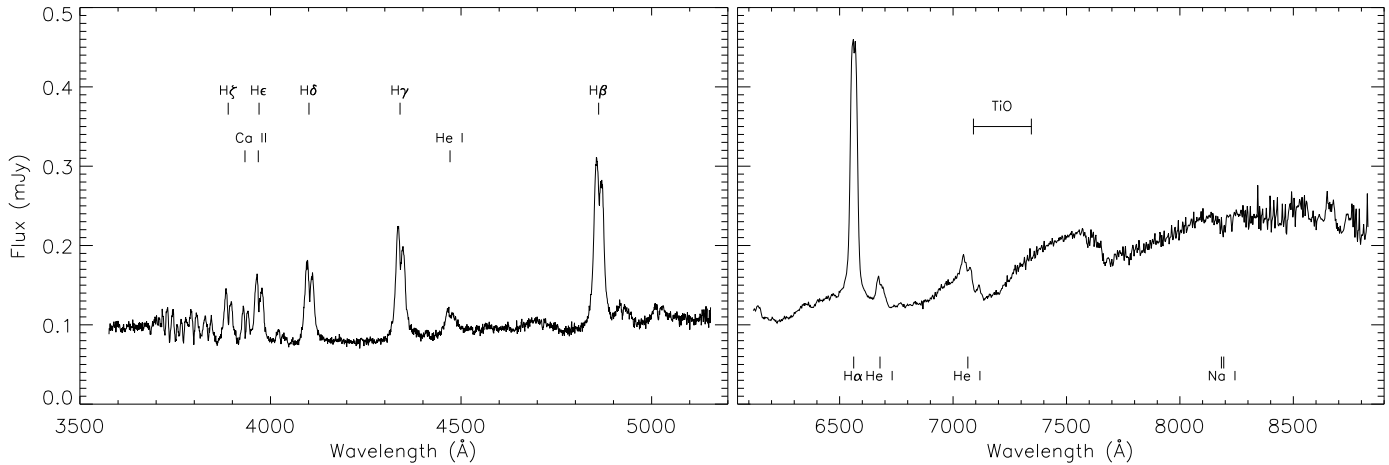}
\caption{\label{fig:whtspec} Flux-calibrated average spectrum of
SDSS\,J1006. Data from the blue arm of ISIS are shown in the left
panel, and from the red arm in the right panel. The most prominent
emission and absorption features are labelled.} \end{figure*}

\subsection{Spectroscopy}                                                           \label{sec:obs:spec}

Spectroscopic observations were obtained in 2008 January, using the ISIS double-beam spectrograph on the William Herschel Telescope (WHT) at La Palma (Table\,\ref{tab:obslog}). For the red arm we used the R316R grating and Marconi CCD binned by factors of 2 (spectral) and 3 (spatial), giving a wavelength range of 6115--8840\,\AA\ at a reciprocal dispersion of 1.85\,\AA\ per binned pixel. For the blue arm we used the R600B grating and EEV12 CCD with the same binning factors \reff{as for the Marconi CCD}, giving a wavelength coverage of 3575--5155\,\AA\ at 0.88\,\AA\ per binned pixel. From measurements of the full widths at half maximum of arc-lamp and night-sky spectral emission lines, we find that this gave resolutions of 3.5\,\AA\ (red arm) and 1.8\,\AA\ (blue arm).

Data reduction was undertaken \reff{by} optimal extraction (\citealt{Horne86pasp}) as implemented in the {\sc pamela}\footnote{{\sc pamela} and {\sc molly} were written by TRM and can be obtained from {\tt http://www.warwick.ac.uk/go/trmarsh}} code \citep{Marsh89pasp}, which also \reff{uses} the {\sc starlink}\footnote{The Starlink software and documentation can be obtained from {\tt http://starlink.jach.hawaii.edu/}} packages {\sc figaro} and {\sc kappa}; further details can be found in \citet{Me+07mn,Me+07mn2}. Copper-neon and copper-argon arc lamp exposures were taken every hour during our observations and the wavelength calibration for each science exposure was linearly interpolated from the two arc observations bracketing it. We removed the telluric lines and flux-calibrated the target spectra using observations of Feige\,110, treating each night separately.

\begin{figure*}  \centering
\includegraphics[width=0.23\textwidth,angle=0]{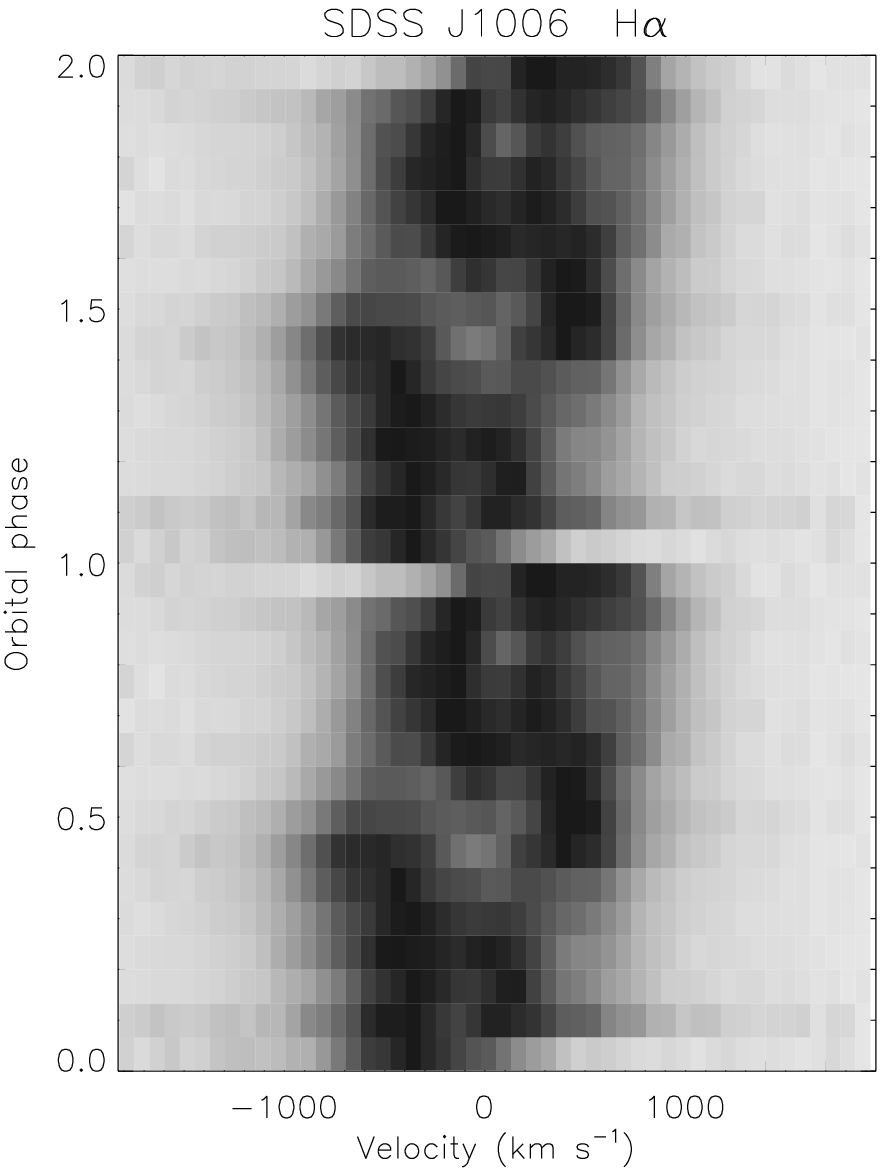} \
\includegraphics[width=0.23\textwidth,angle=0]{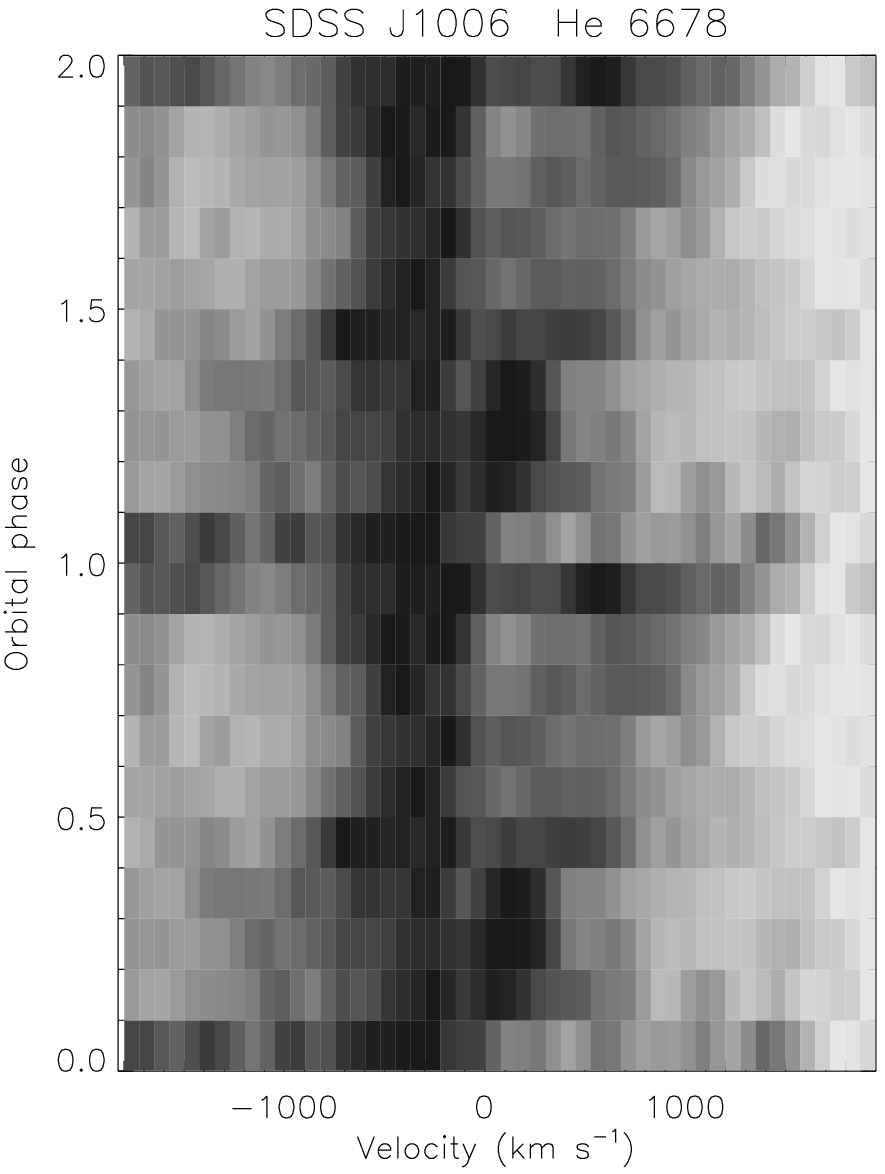} \
\includegraphics[width=0.23\textwidth,angle=0]{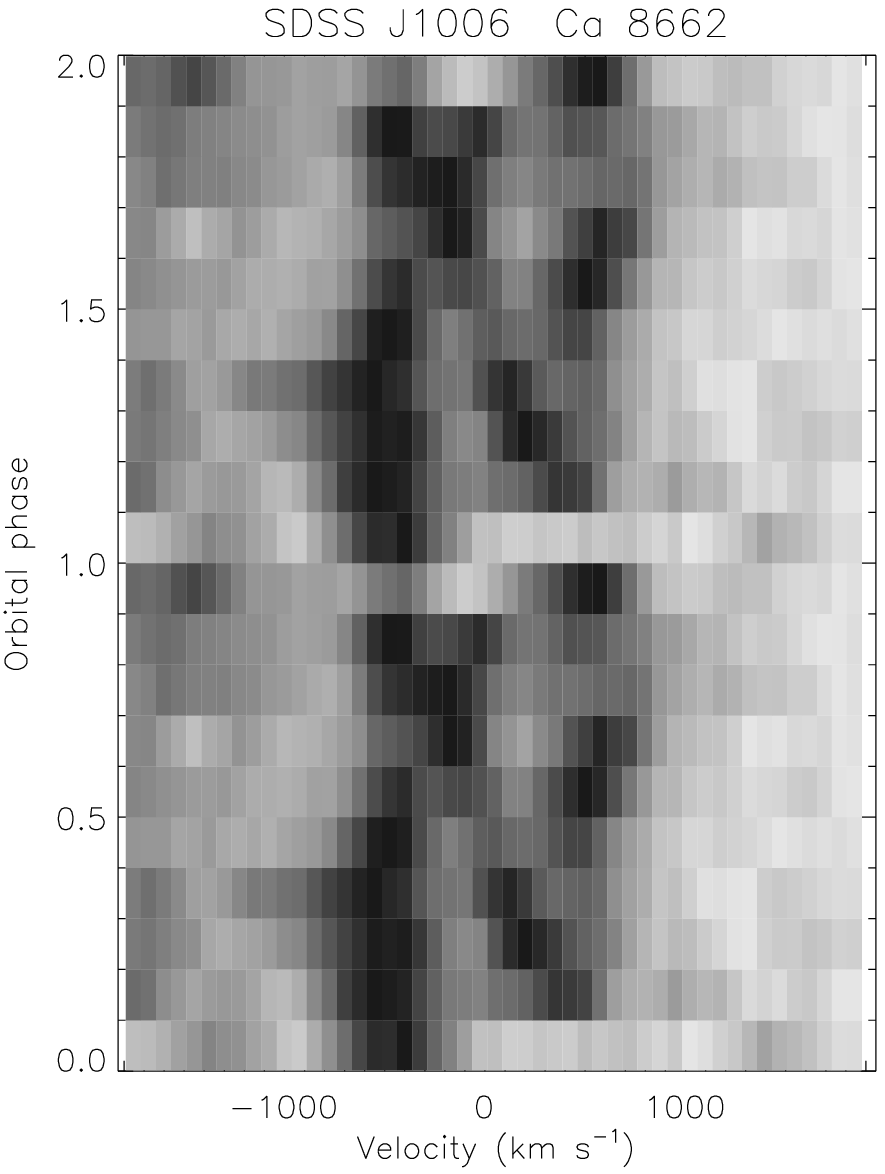} \
\includegraphics[width=0.23\textwidth,angle=0]{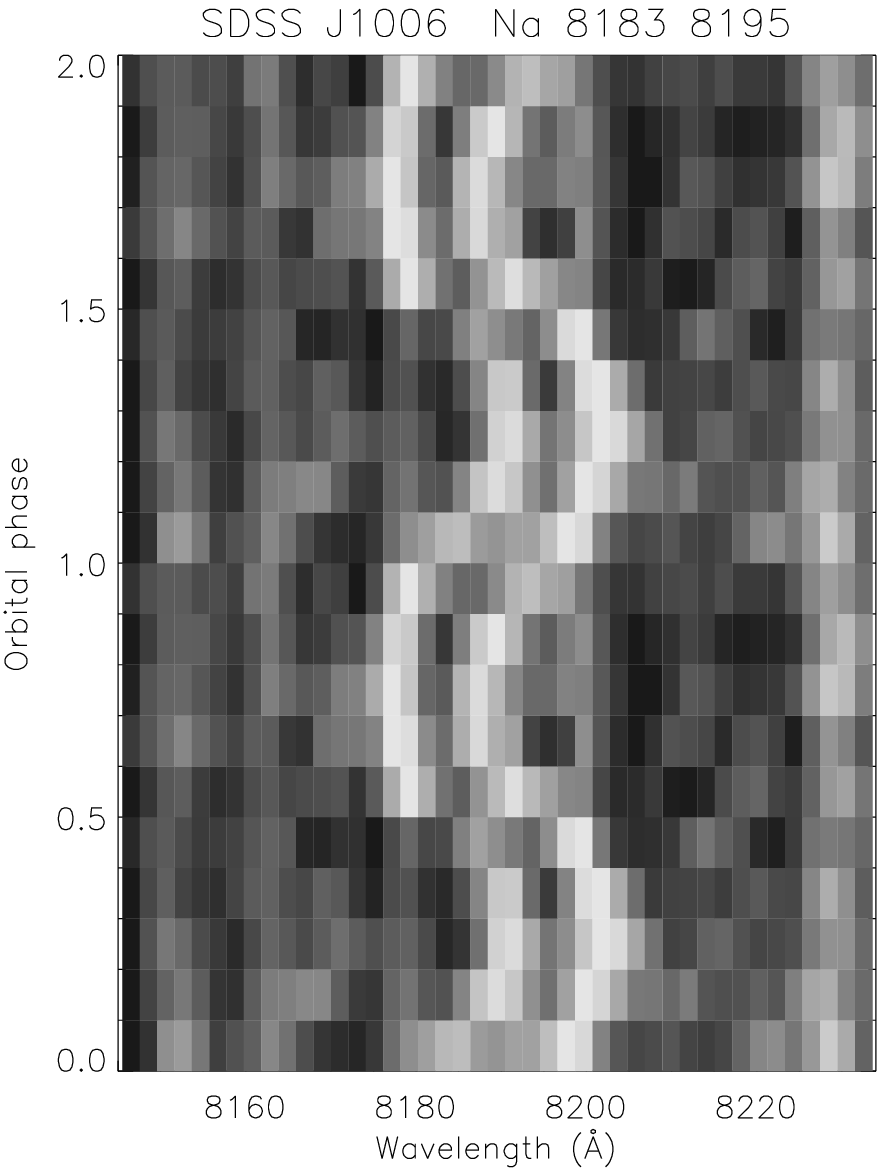} \\
\caption{\label{fig:trailed} Greyscale plot of the continuum-normalised
and phase-binned trailed spectra of SDSS\,J1006. From left to right the
panels show H$\alpha$, \ion{He}{I} 6678\,\AA\ and \ion{Ca}{II} 8662\,\AA\
emission, and \ion{Na}{I} 8183 and 8194\,\AA\ absorption. The plots for
\ion{He}{I} and \ion{Ca}{II} have been smoothed in wavelength with a
Savitsky-Golay filter for display purposes.} \end{figure*}

The averaged WHT spectra are shown in Fig.\,\ref{fig:whtspec}. Trailed greyscale plots of the phase-binned spectra are shown in Fig.\,\ref{fig:trailed}, for the H$\alpha$, \ion{He}{I} 6678\,\AA\ and \ion{Ca}{II} 8662\,\AA\ emission lines, and \ion{Na}{I} 8183 and 8194\,\AA\ absorption lines, and will be discussed in Section\,\ref{sec:analysis}.

\subsection{Photometry}                                                                   \label{sec:obs:phot}

Time-series photometry of SDSS\,J1006 was obtained in service mode using two telescopes equipped with imaging spectrographs: the Nordic Optical Telescope (NOT) with ALFOSC, and the Calar Alto (CAHA) 2.2\,m telescope with CAFOS. For the NOT observations we used the No.\,92 filter, which has a wide-$V$ passband with points of half transmission at approximately 4400 and 7000\,\AA. The CAHA observations were made unfiltered to maximise throughput. The CCDs were mostly binned and windowed to reduce readout time, and short exposure times were used to maximise the cadence of the observations.

The 2008 December observations obtained with the NOT were reduced with the pipeline described by \citet{Me+09mn2,Me+09mn}, which uses an {\sc idl} implementation of {\sc daophot} to perform aperture photometry. The remaining photometric data were reduced using the pipeline described by \citet{gansicke+04aa}, which performs bias and flat-field corrections within {\sc midas}\footnote{\tt http://www.eso.org/projects/esomidas/} and aperture photometry with the {\sc SExtractor} package \citep{BertinArnouts96aas}. Instrumental differential magnitudes were converted into $V$-band apparent magnitudes, using the SDSS $ugriz$ apparent magnitudes of several comparison stars and the transformation equations given by Lupton\footnote{The $ugriz-BVRI$ transformation equations are attributed to ``Lupton (2005)'' but appear to be unpublished. They can be found at {\tt http://www.sdss.org/dr6/algorithms/sdssUBVRITransform.html}}.

\begin{figure} \centering
\includegraphics[width=0.48\textwidth,angle=0]{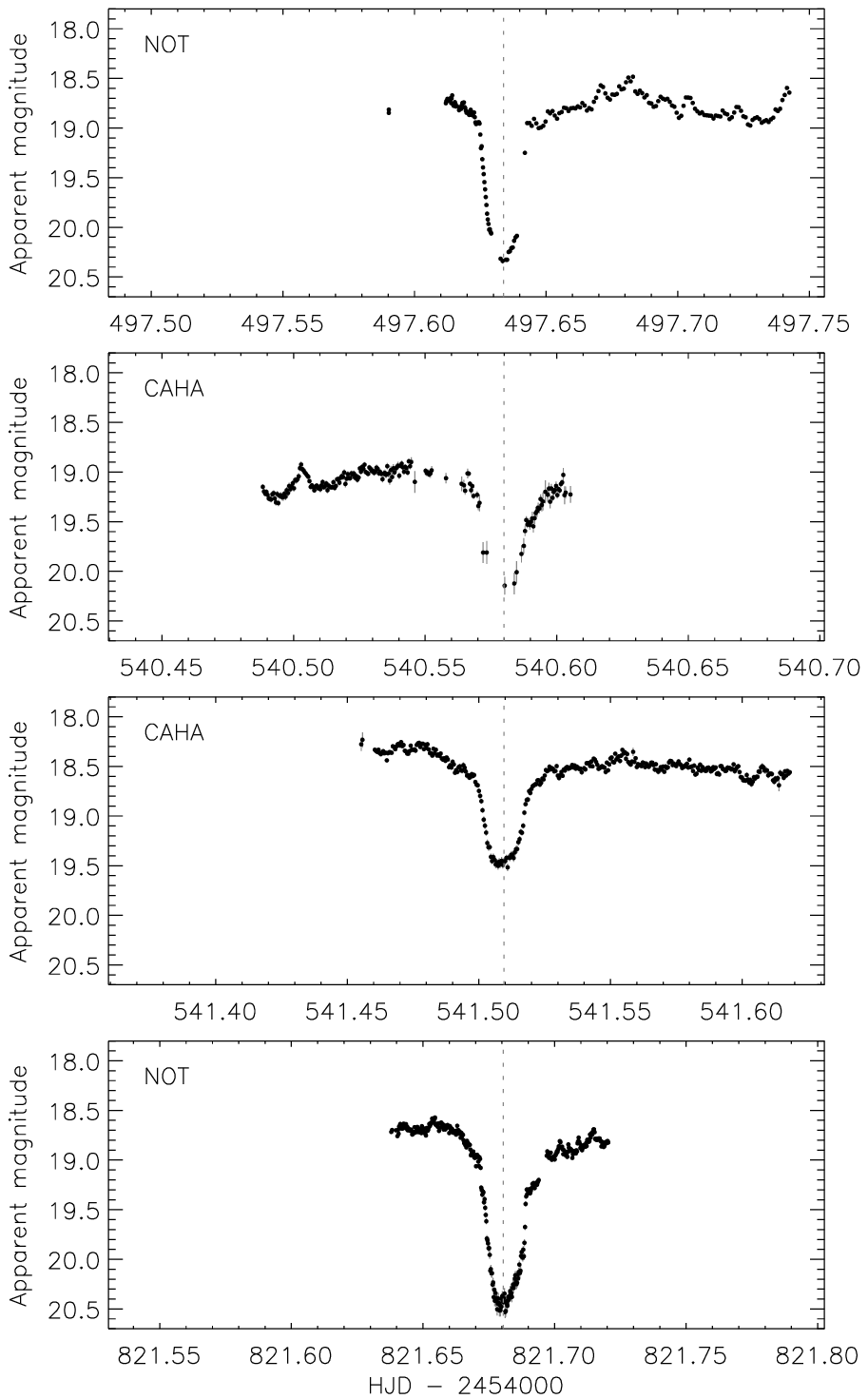} \\
\caption{\label{fig:lcall} Plot of the four light curves obtained
covering eclipses of SDSS\,J1006. The eclipse midpoints have been
aligned on the panels.} \end{figure}

The light curves are plotted in Fig.\,\ref{fig:lcall} with the measured eclipse midpoints indicated. It is apparent from this plot that the depth of the eclipse is dependent on the wavelength of observation: the Calar Alto data were unfiltered, so are more affected by the light of the secondary star and thus show shallower eclipses.

%%%%%%%%%%%%%%%%%%%%%%%%%%%%%%%%%%%%%%%%%%%%%%%%%%%%%%%%%%%%%%%%%%%%%%%%%%%%%%%%%%%%%%%%%%%%%%%%%%%%%%%%%%%%%%

\section{Analysis}                                                                   \label{sec:analysis}

\subsection{Orbital ephemeris}                                             \label{sec:porb}

Our first observations of SDSS\,J1006 were spectroscopic. Radial velocities (RVs) measured from the emission lines (see below) clearly showed anomalies due to three eclipses, on an unambiguous period \reff{of} 267.9\,min. The resulting preliminary ephemeris was sufficiently accurate for us to photometrically observe eclipses, on which more precise period measurements could be based.

For each of the four eclipses, a mirror-image of the light curve was shifted until the two representations of the central parts of the eclipse were in the best possible agreement. The time defining the axis of reflection was taken as the midpoint and uncertainties were estimated based on how far this could be shifted before the agreement was clearly poorer. We have fitted a linear ephemeris to these times of minimum light, finding
\[ {\rm Min\,I (HJD}) = 2454540.57968 (40) + 0.18591324 (42) \times E \]
where $E$ is the cycle number and the parenthesised quantities indicate the uncertainty in the last digit of the preceding number. This corresponds to an orbital period of $267.71507 \pm 0.00060$\,min. The measured times of minimum light and the observed minus calculated values are given in Table\,\ref{tab:eclipses}. All phases in this work are calculated using this ephemeris.

\begin{table} \begin{center}
\caption{\label{tab:eclipses} Times of eclipse for
SDSS\,J1006 and the residuals with respect to the
linear ephemeris given in Section\,\ref{sec:porb}.}
\begin{tabular}{l r@{\,$\pm$\,}l r} \hline
Cycle & \mc{Time of eclipse (HJD)} & Residual (d) \\
\hline
 -231 &  2454497.6338  &  0.0010   &    0.0001    \\
    0 &  2454540.5802  &  0.0010   &    0.0005    \\
    5 &  2454541.5091  &  0.0005   & $-$0.0002    \\
 1512 &  2454821.6805  &  0.0005   &    0.0000    \\
\hline \end{tabular} \end{center} \end{table}

\subsection{Emission-line radial velocities}                                                 \label{sec:em}

\begin{figure} \centering
\includegraphics[width=0.48\textwidth,angle=0]{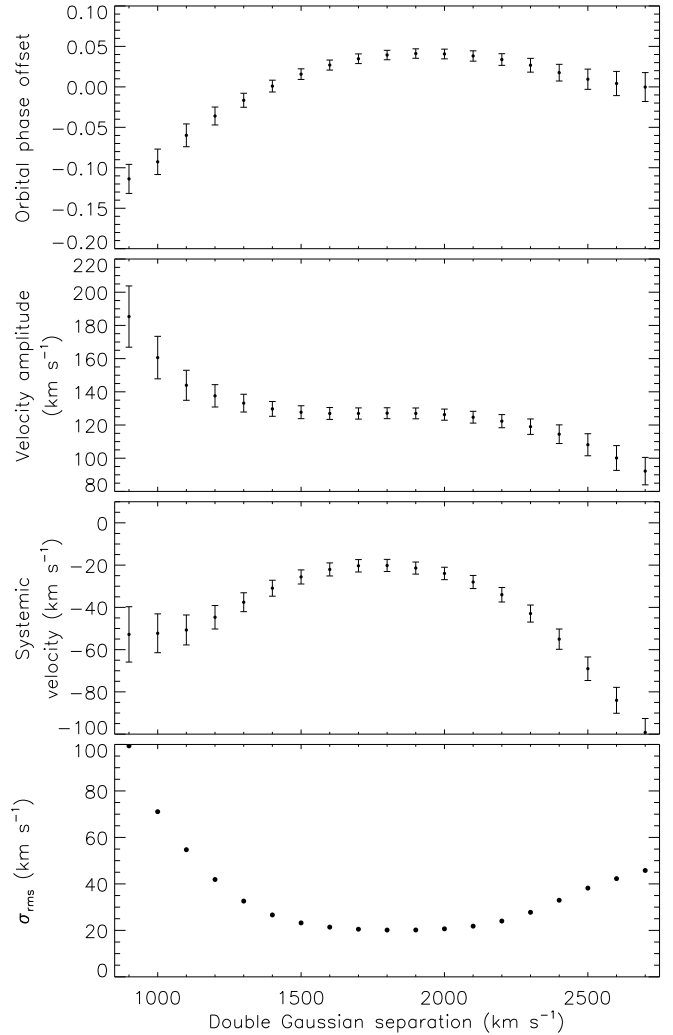} \\
\caption{\label{fig:diagnostic} Diagnostic diagram showing the
variation of the best-fitting spectroscopic orbital parameters
for RVs measured with a range of separations using the double
Gaussian function. $\sigma_{\rm rms}$ denotes the scatter of
the RV measurements around the fitted orbit.} \end{figure}

% \begin{table} \centering \caption{\label{tab:rvorbit1} Best-fitting
% emission-line spectroscopic orbit found using {\sc sbop}. The
% reference time is a time of maximum negative rate of change of
% RV. The uncertainties include random and systematic contributions.}
% \begin{tabular}{l c}\hline
% Reference time (HJD)                & 2454497.6411 $\pm$ 0.0023  \\
% Orbital period (d)                  & 0.18591324 (fixed)         \\
% Eccentricity                        & 0.0 (fixed)                \\
% Velocity amplitude $K_1$ (\kms)     & 127.1 $\pm$ 4.6            \\
% Systemic velocity (\kms)            & $-$20.2 $\pm$ 6.0          \\
% $\sigma_{\rm rms}$ (\kms)           & 20.1                       \\
% \hline \end{tabular} \end{table}

\begin{table} \centering \caption{\label{tab:rvorbit1}\label{tab:rvorbit2}
Best-fitting spectroscopic orbits found using {\sc sbop}. The reference
times are time of maximum negative rate of change of RV. The uncertainties
include both random and systematic contributions.}
\begin{tabular}{l c}\hline
Orbital period (d)              & 0.18591324 (fixed)        \\
Eccentricity                    & 0.0 (fixed)               \\
\hline
\multicolumn{2}{l}{\it Emission-line radial velocities:}    \\
Reference time (HJD)            & 2454497.6411 $\pm$ 0.0023 \\
Velocity amplitude $K_1$ (\kms) & 127.1 $\pm$ 4.6           \\
Systemic velocity (\kms)        & $-$20.2 $\pm$ 6.0         \\
$\sigma_{\rm rms}$ (\kms)       & 20.1                      \\
\hline
\multicolumn{2}{l}{\it Absorption-line radial velocities:}  \\
Reference time (HJD)            & 2454497.63381 (fixed)     \\
Measured $K_2$ (\kms)           & 276 $\pm$ 7               \\
Corrected $K_{\rm MD}$ (\kms)   & 258 $\pm$ 12              \\
Systemic velocity (\kms)        & $-$20 $\pm$ 10            \\
$\sigma_{\rm rms}$ (\kms)       & 20.8                      \\
\hline \end{tabular} \end{table}

\begin{figure} \centering
\includegraphics[width=0.48\textwidth,angle=0]{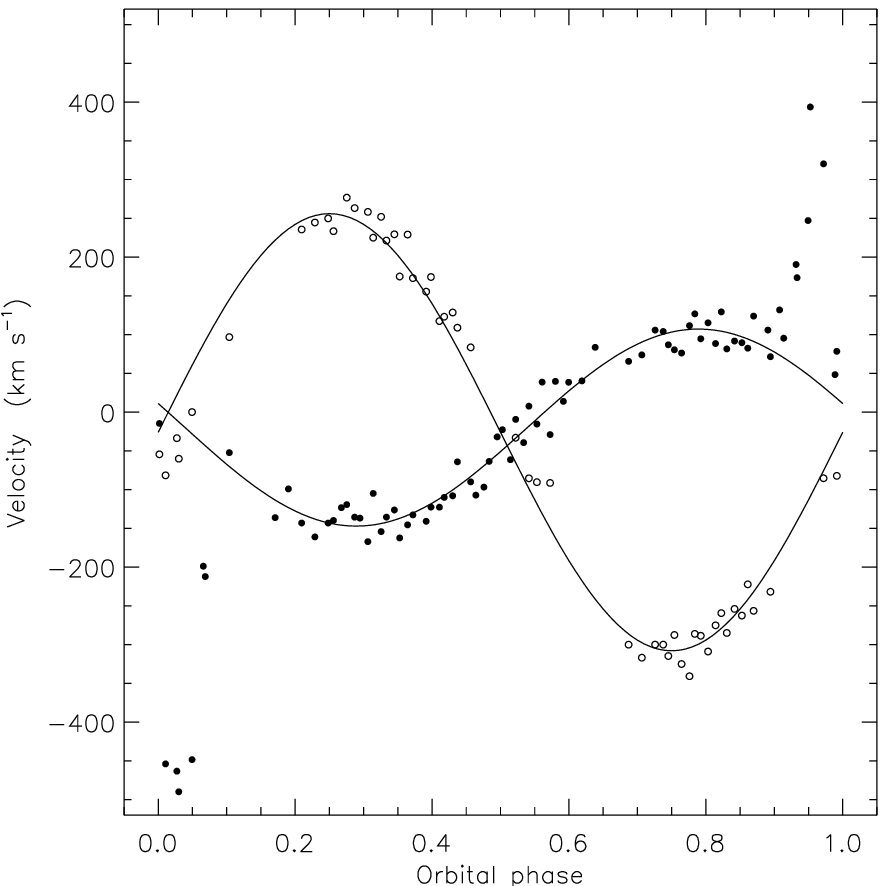}\\
\caption{\label{fig:rvplot} The measured RVs (circles) and the spectroscopic
orbits fitted to them (solid curves). The emission-line RVs (filled circles)
were calculated using $\xi = 1800$\kms. The measurements at phases 0.9--0.1
are affected by the eclipse of the accretion disc and were not included in the
fit. The absorption-line RVs (open circles) were obtained by cross-correlation
against an M4 template spectrum, and include only those spectra which yielded
a reliable cross-correlation function.} \end{figure}

The spectrum of SDSS\,J1006 (Fig.\,\ref{fig:whtspec}) shows strong emission at the wavelengths of the hydrogen Balmer lines and some helium lines. These emission lines are produced by the accretion disc which surrounds the WD, so variations in their velocity hold information on the motion of the WD itself. However, spectroscopic studies of CVs often show a phase difference between the RV variation of emission lines and the orbital phases measured using other methods \citep{Thorstensen00pasp,Thoroughgood+05mn,Unda++06mn,Steeghs+07apj}. This casts doubt on whether emission lines are good indicators of the motion of WDs in CVs, and due to this we did not use emission-line RVs in calculating the physical properties of SDSS\,J1006.

We measured RVs from the H$\alpha$ emission, which is the strongest emission line, using the double-Gaussian method \citep{SchneiderYoung80apj} as implemented in {\sc molly}. The full width half maximum of the Gaussian functions was set to 300\kms, which is a good compromise between resolving emission-line features and minimising the random noise in the RV measurements. The separation of the two Gaussians, $\xi$, was varied from 800 to 3000\kms\ in jumps of 100\kms. For each value of $\xi$ a spectroscopic orbit was fitted to the measured RVs using the {\sc sbop}\footnote{Spectroscopic Binary Orbit Program, written by P.\ B.\ Etzel, \\ {\tt http://mintaka.sdsu.edu/faculty/etzel/}} code, which we find gives reliable error estimates for the optimised parameters \citep{Me+05mn}. The orbital period was fixed at the ephemeris value (Section\,\ref{sec:porb}), a circular orbit was assumed, and the phase zeropoint was included as a fitted parameter. RVs between phases 0.9 and 0.1 were rejected as they are affected by the eclipse of the accretion disc by the secondary star.

We have constructed a diagnostic diagram \citep{Shafter83apj,Shafter++86apj} for SDSS\,J1006 (Fig.\,\ref{fig:diagnostic}), which shows that the properties of the spectroscopic orbit change only slowly for $\xi = 1600$--2200\kms, and that the lowest scatters in the residuals ($\sigma_{\rm rms}$) occurs for $\xi = 1700$--2000\kms. %We would expect that the most reliable RVs will be found for larger values of $\xi$ (within the limits set by the amount of observational noise): these are sensitive to the flux in the wings of the emission line, which comes from the higher-velocity parts of the accretion disc which are close to the WD.
The offset between the orbital phase and the phase of greatest negative change in the RVs is only about 0.04 for these separations, which indicates that the emission-line RVs might \reff{trace the motion of the WD with reasonable accurately}. We have adopted the spectroscopic orbit for $\xi = 1800$\kms, which gives the lowest $\sigma_{\rm rms}$, and these quantities are given in Table\,\ref{tab:rvorbit1}. The RVs and best fit are shown in Fig.\,\ref{fig:rvplot}. Our error estimates include the standard errors given by {\sc sbop}, plus a contribution to account for variations between the solutions for $\xi = 1500$--2200\kms\ (where the $\sigma_{\rm rms}$ \reff{values} are the lowest). We also calculated a diagnostic diagram for the H$\beta$ emission line, which yielded similar results but a greater scatter due to the weaker emission-line flux.

\subsection{Absorption-line radial velocities}                                                  \label{sec:abs}

% \begin{table} \centering \caption{\label{tab:rvorbit2}
% Best-fitting absorption-line spectroscopic orbit found using
% {\sc sbop}. The reference time is a time of eclipse midpoint.}
% \begin{tabular}{l c}\hline
% Reference time (HJD)          & 2454497.63381 (fixed) \\
% Orbital period (d)            & 0.18591324 (fixed)    \\
% Eccentricity                  & 0.0 (fixed)           \\
% Measured $K_2$ (\kms)         & 276 $\pm$ 7           \\
% Corrected $K_{\rm MD}$ (\kms) & 263 $\pm$ 12          \\
% Systemic velocity (\kms)      & $-$20 $\pm$ 10        \\
% $\sigma_{\rm rms}$ (\kms)     & 20.8                  \\
% \hline \end{tabular} \end{table}

The secondary component of SDSS\,J1006 is clearly visible in our red-arm WHT/ISIS spectra, but very few features can be seen by the naked eye to vary in velocity, due to the modest signal-to-noise ratio of individual spectra. This velocity variation plays a vital role in constraining the properties of the system, so we have used two methods to tease out the absorption-line velocity amplitude.

Firstly, the observed spectra were augmented with a set of template M dwarf spectra from the SDSS, then velocity-binned and subjected to a cross-correlation analysis. The orbital ephemeris was fixed to the numbers in Section\,\ref{sec:porb} after verifying that this does not cause a significant change in the results. The cross-correlation functions were examined interactively and measured for velocity if they contained a clear peak from the secondary star, and the resulting RVs were fitted with a circular orbit using {\sc sbop}. We did this for many different spectral regions and template spectra, finding that the resulting velocity amplitudes were always in the interval 270--282\kms. For illustration, in Fig.\,\ref{fig:rvplot} we plot the absorption-line RVs found using an M4 spectral template and the full red-arm wavelength interval (with the H$\alpha$ and helium emission lines masked \reff{out}).

The second method is designed to cope well with spectra of a low signal-to-noise ratio, and to target the strongest spectral features observed to come from the secondary star. Using the {\sc mgfit} routine in {\sc molly} we fitted a double Gaussian function plus spectroscopic orbit to the sodium doublet at 8183.3 and 8194.8\,\AA. All 79 spectra were fitted simultaneously, yielding a direct measurement of the velocity amplitude: $K_2 = 275.8 \pm 3.6$\kms. Fig.\,\ref{fig:trailed} shows the phase-binned and trailed spectra of SDSS\,J1006 in the region of the Na doublet. We have been unable to completely remove the effects of telluric absorption from our spectra, so have also performed fits with extra Gaussians added to account for the residual absorption. We find that our $K_2$ measurement is not significantly affected.

Given the good agreement between the two methods, we adopt a value of $K_2 = 276 \pm 7$\kms, where the error estimate accounts for both the random errors and the variation in results from different analysis techniques (Table\,\ref{tab:rvorbit2}). The two methods agree well on the value of $K_2$ but produce slightly discrepant systemic velocity measurements. This is likely due to difficulties in placing the continuum, due to the complex spectrum of the secondary star. We adopt a value of $V_\gamma = -20 \pm 10$\kms, which encompasses most of the systemic velocities found during our analysis. A better measurement of $V_\gamma$ will require further data.

\subsubsection{$K$-correction for the absorption-line velocities}                            \label{sec:abs:k}

\begin{figure} \centering
\includegraphics[width=0.48\textwidth,angle=0]{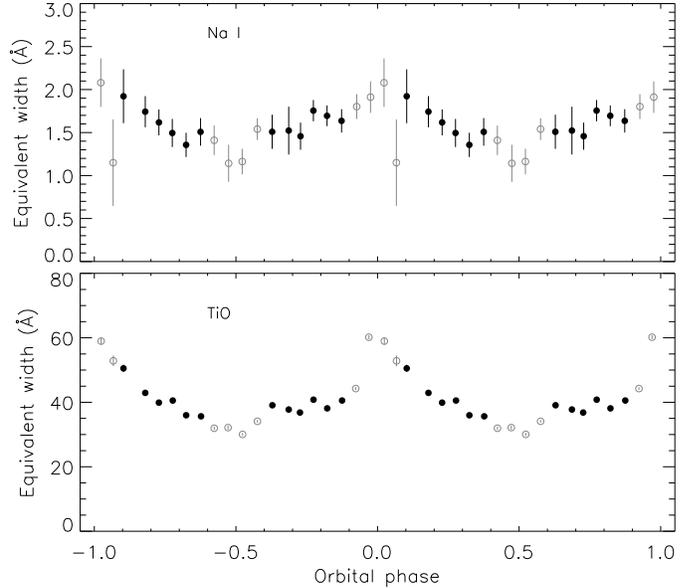}\\
\caption{\label{fig:ew} Variation of the equivalent widths of the
Na doublet and TiO molecular band with orbital phase. The spectra
were combined into 20 phase bins prior to measurement. The black
filled circles represent measurements outside primary and secondary
eclipses, and the grey open circles those within eclipse. The
wavelength intervals over which the equivalent widths were measured
were 8180--8200\,\AA\ for Na and 7090--7345\,\AA\ for TiO.} \end{figure}

Our measured $K_2$ cannot be assumed to represent the motion of the centre of mass of the secondary star, $K_{\rm MD}$, due to irradiation of the inner hemisphere by the WD and accretion disc. The irradiated surface has a lower vertical temperature gradient and thus weaker absorption lines. RV measurements from these lines are therefore skewed towards the motion of the outward-facing part of the star, causing $K_2$ to overestimate $K_{\rm MD}$ \citep{Hessman+84apj,WadeHorne88apj,Billington++96mn}.

To estimate the correction $\Delta K = K_2 - K_{\rm MD}$ we have measured the equivalent widths of absorption features arising from the secondary star, as a function of orbital phase. The wavelength scales of the spectra were moved to shift out the motion of the star, and the spectra were then rectified to a continuum level of 1 and binned into twenty phase intervals. The resulting plots (Fig.\,\ref{fig:ew}) show that the equivalent widths vary by approximately 30\% outside eclipse. Extrapolating to the phase of secondary mid-eclipse and considering the errors on this approach and on the equivalent width measurements, we find a total variation in equivalent width (and thus in the light from the secondary star) of $35 \pm 15$\%.

\citet{WadeHorne88apj} obtained the expression \[ \Delta K = f \frac{R_{\rm MD}}{a_{\rm MD}} K_{\rm MD} \] where $f$ is the size of the displacement as a fraction of $R_{\rm MD}$, the radius of the secondary star, and $a_{\rm MD}$ is the semimajor axis of the orbit of this component. The light curve analysis (see below) gives a mass ratio of $q = 0.51 \pm 0.10$, which results in a secondary star radius of $R_{\rm MD} = (0.49 \pm 0.06) a_{\rm MD}$. The largest value of $f$ is $4/3\pi \approx 0.42$ and occurs when the spectral lines are totally quenched on the irradiated hemisphere of the star. We therefore adopted $f = 0.42 \times (0.35 \pm 0.15) = 0.15 \pm 0.06$, resulting in a $K$-correction of $\Delta K = (0.07 \pm 0.04) K_{\rm MD}$. Armed with this correction we have determined the velocity amplitude of the centre of mass of the secondary star to be $K_{\rm MD} = 258 \pm 12$\reff{\kms} (Table\,\ref{tab:rvorbit2}).

\subsection{Doppler tomography}                                                                 \label{sec:doppler}

\begin{figure*}  \centering
\includegraphics[width=0.32\textwidth,angle=0]{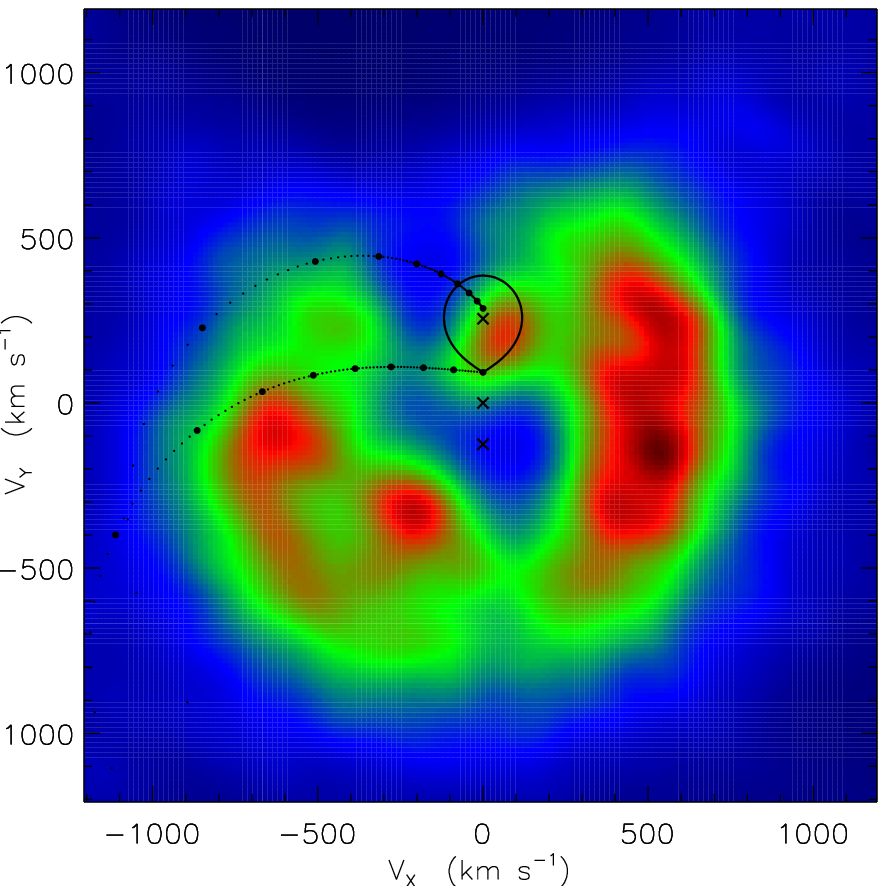} \ \
\includegraphics[width=0.32\textwidth,angle=0]{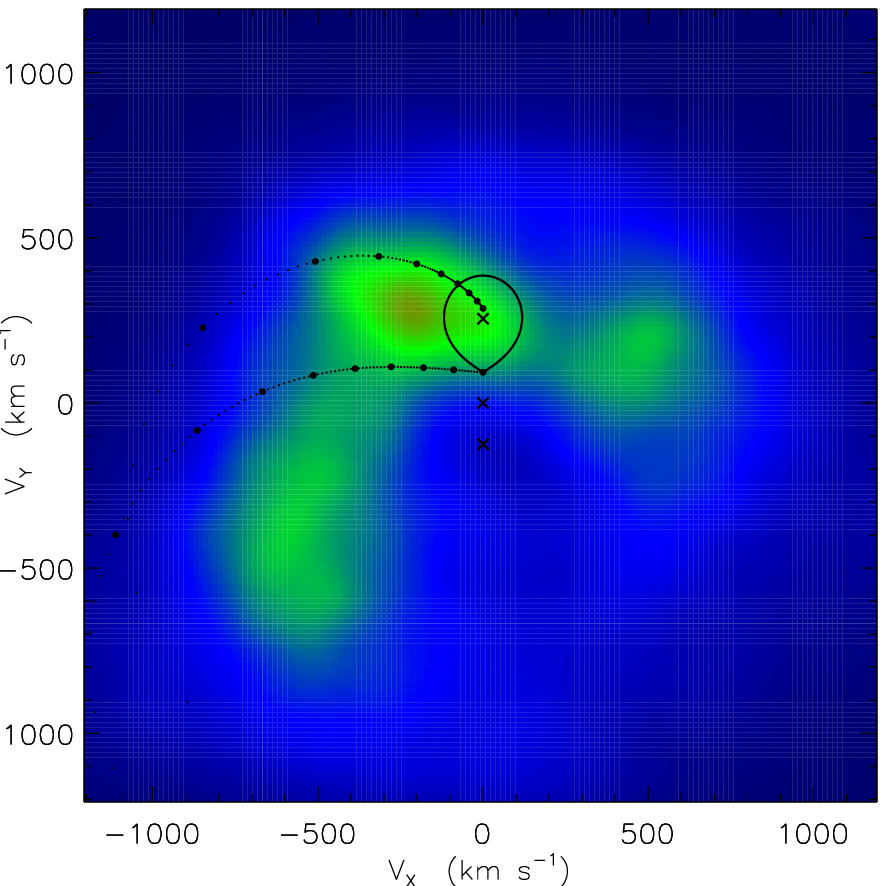} \ \
\includegraphics[width=0.32\textwidth,angle=0]{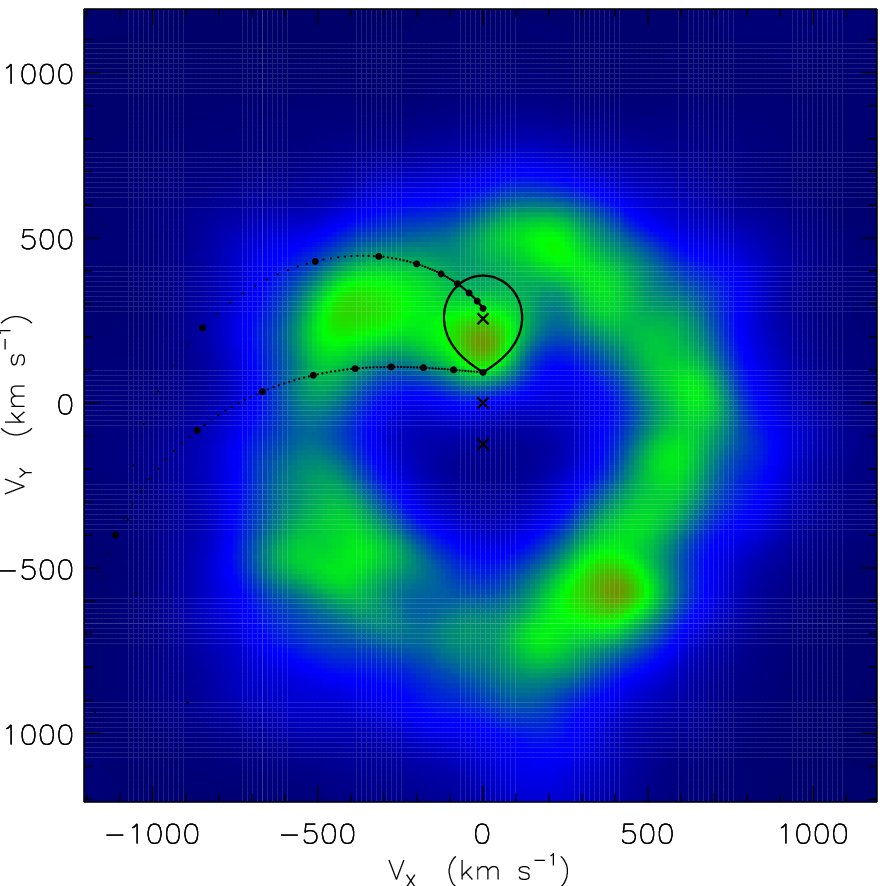} \\
\caption{\label{fig:dopmap} Doppler maps of H$\alpha$ (left),
\ion{He}{I} 6678\,\AA\ (centre) and \ion{Ca}{II} 8662\,\AA\ (right).
Assuming $K_1 = 127$\kms\ and $K_2 = 258$\kms, the Roche lobe of the
secondary is shown with a solid line, the centres of mass of the system
and of the two stars are shown by crosses, and the velocity of the
accretion stream and the Keplerian velocity of the accretion disc are
indicated by dots with a constant spacing in position. The orientation
of the maps has been set using the eclipse ephemeris.} \end{figure*}

To investigate the properties of the accretion disc of SDSS\,J1006 we have constructed Doppler maps of several of the emission lines using the maximum entropy method \citep{MarshHorne88mn}. The maps are shown in Fig.\,\ref{fig:dopmap} and phase-binned and trailed plots of the emission lines are shown in Fig.\,\ref{fig:trailed}. The $\chi^2$ value for the Doppler maps were chosen to be marginally larger than the values for which noise features start to be visible, and the orientation of the maps was specified using the eclipse ephemeris. Overlaid on the Doppler maps are interpretations of the system properties, adopting $K_{\rm WD} = 127$\kms\ and $K_{\rm MD} = 258$\kms.

The Doppler maps for the Balmer emission lines (see the H$\alpha$ map in Fig.\,\ref{fig:dopmap}) have an unusual wide double-lobed structure. The H$\alpha$ map also shows emission attributable to the secondary star, although this is oddly offset from the line of centres of the system. The shape of the accretion disc and the offset of the secondary star emission may be artefacts of the breakdown of an important assumption of Doppler tomography: that emitting regions are optically thin.

Doppler maps of the \ion{He}{I} emission lines show weak and diffuse emission in the region of the bright spot, which is where the accretion stream from secondary star encounters the edge of the accretion disc. The bright spot is not a major source of \ion{He}{I} emission, but very little else is seen in the \ion{He}{I} maps.

We have also constructed a Doppler map of the \ion{Ca}{II} 8662\,\AA\ emission, which is the line of the calcium triplet which is least affected by night-sky emission and telluric lines. The map (Fig.\,\ref{fig:dopmap}) shows a circular accretion disc feature and clear emission arising from the secondary star. The latter emission can be seen describing a S-wave in trailed spectra (Fig.\,\ref{fig:trailed}), and its position in the map supports our measurement of the velocity amplitude of this star.

\subsection{Light curve modelling}                                                              \label{sec:lc}

\begin{table} \centering
\caption{\label{tab:lcfit} Results of the light curve modelling process. The
formal uncertainties come from the MCMC analysis and the adoped uncertainties
are increased to account for several additional sources of uncertainty. Several
parameters are given in units of $a$, the \reff{orbital} semimajor axis.}
\setlength{\tabcolsep}{4pt}
\begin{tabular}{lccc}\hline
Quantity                      & Value           & Formal          & Adopted         \\
\                             &                 & uncertainty     & uncertainty     \\
\hline
Reference time (HJD)          & 2454821.68051   & 0.00002         & 0.0002          \\
Orbital period (d)            & 0.18591324      &         \mc{(fixed)}              \\
Orbital inclination ($^\circ$)& 81.3            & 0.8             & 2.0             \\
Mass ratio                    & 0.51            & 0.04            & 0.08            \\
Disc radius ($a$)             & 0.189           & 0.015           & 0.05            \\
White dwarf radius ($a$)      & 0.0110          & 0.0013          & 0.003           \\
Secondary star radius ($a$)   & 0.322           & 0.006           & 0.010           \\
\hline \end{tabular} \end{table}

\begin{figure} \centering
\includegraphics[width=0.48\textwidth,angle=0]{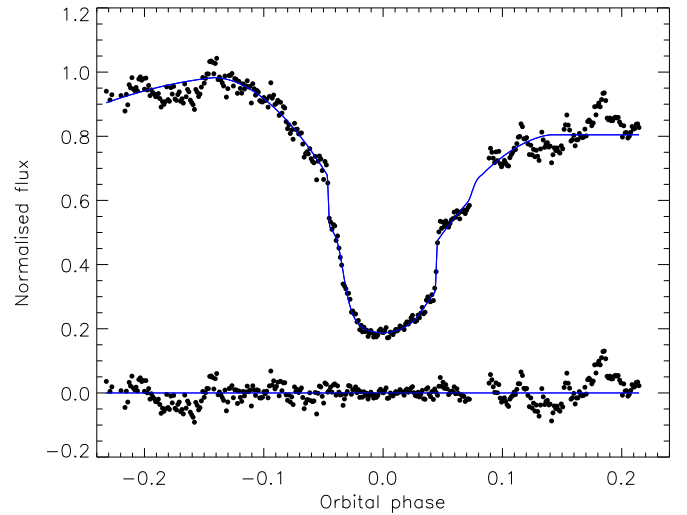}
\caption{\label{fig:lcfit} Light curve of SDSS\,J1006 from the 2008
December observations (points) compared to the best fit found using
{\sc lcurve} (solid curve). The residuals of the fit are plotted at
the base of the figure.} \end{figure}

Our light curves show deep eclipses due to obscuration of the \reff{WD and} accretion disc by the secondary star. To obtain constraints on the system properties, the best dataset (2008 December) was compared to synthetic light curves created using the {\sc lcurve} code written by TRM (see \citealt{Pyrzas+09mn}). This uses grids of points to model the WD as a sphere, the secondary star using Roche geometry, a flat circular \reff{accretion} disc, and an exponentially decreasing bright spot. The best fit to the observed data was obtained with a combination of the downhill simplex and Levenberg-Marquardt algorithms \citep{Press+92book}. The outside-eclipse data show strong stochastic variation (termed flickering; see \citealt{Bruch92aa} and \citealt{Bruch00aa}) arising from the mass-transfer process in SDSS\,J1006. We have down-weighted data outside the phase interval [$-0.07,0.08$] by a factor of three, to limit their influence on the fit.

The small gap in the light curve at HJD 2454821.69 is unfortunate, as the egress of the bright spot occurs somewhere during this time. We find two main families of good fits to the light curve corresponding to different WD radii and mass ratios: the first family is in the region of $R_{\rm WD} = 0.011 a$ and $q=0.51$ and is our preferred solution. The second centres on $R_{\rm WD} = 0.023 a$, which is unphysically large, and $q=0.60$. After extensive exploration of the parameter space we adopt the first solution but increase the errorbars of the light curve parameters to include the full range of plausible solutions which we found.

Internal parameter errors were determined by $10^5$ Markov Chain Monte Carlo (MCMC) simulations. For these simulations we accepted a certain fraction of random jumps in parameter values and evaluated how they changed the quality of the fit. After the simulations showed reasonable convergence, the errors and covariances could be computed from examining the parameters from the accepted jumps. We rejected typically the first 10\% of values to avoid a dependence on the initial parameter values. This gives a more realistic view of the parameter uncertainties compared to the values computed simply from the analytic errors alone, an aspect which is particularly important given the correlated noise due to flickering.

The results of the light curve modelling process are given in Table\,\ref{tab:lcfit} and the best-fitting model is compared to the data in Fig.\,\ref{fig:lcfit}. The radii of the stars and accretion disc are given in units of the orbital semimajor axis, $a$. The uncertainties yielded by the MCMC analysis still do not fully take into account the flickering or the range of plausible solutions we found. We have increased the uncertainties to include the full range of reasonable trial solutions we found (Table\,\ref{tab:lcfit}), and regard the results as conservative. A substantial improvement will require high-speed photometry of several eclipses of the SDSS\,J1006 system.

%%%%%%%%%%%%%%%%%%%%%%%%%%%%%%%%%%%%%%%%%%%%%%%%%%%%%%%%%%%%%%%%%%%%%%%%%%%%%%%%%%%%%%%%%%%%%%%%%%%%%%%%%%%%%%

\section{The physical properties of SDSS\,J1006}

\begin{table} \centering
\caption{\label{tab:final} Physical properties of the stellar components of SDSS\,J1006.}
\begin{tabular}{l c c}\hline
Quantity                    & White dwarf              & M dwarf                  \\
\hline
Semimajor axis (\Rsun)      & \multicolumn{2}{c}{$1.45 \pm 0.10$}                 \\
Mass (\Msun)                & $0.78 \pm 0.12$          & $0.40 \pm 0.10$          \\
Radius (\Rsun)              & $0.016 \pm 0.006$        & $0.466 \pm 0.036$        \\
\logg\ [cm\,s$^{-2}$]       & $7.93 \pm 0.33$          & $4.701 \pm 0.079$        \\
\hline \end{tabular} \end{table}

The spectroscopic and eclipsing characteristics of SDSS\,J1006 allow the determination of the masses and radii of the WD and its low-mass companion. From measurements of the times of mid-eclipse we have obtained an accurate orbital period of 0.18591324(41)\,d. From the infrared sodium doublet we have measured the velocity amplitude $K_2 = 276 \pm 7$\kms. A correction for irradiation effects leads to the secondary star velocity amplitude $K_{\rm MD} = 258 \pm 12$\kms. From modelling the eclipse morphology of SDSS\,J1006 we have found an orbital inclination of $i = 81.3^\circ \pm 2.0^\circ$ and a mass ratio of $q = 0.51 \pm 0.08$.

Combining these results \reff{yields} the masses and radii of the WD and secondary star in SDSS\,J1006 (Table\,\ref{tab:final}). The mass of the former, 0.78\Msun, is higher than the average for single WDs, in agreement with previous results for CVs \citep{SmithDhillon98mn,Littlefair+08mn}. Its radius is consistent (within its large uncertainty) with the theoretical mass--radius relationship for a $\Teff = 15\,000$\,K WD \citep{Bergeron++95apj}.

The secondary star has a mass of 0.40\Msun\ and a radius of 0.47\Rsun, which is distended compared to a normal object -- a mass--radius relation based on detached eclipsing binary star systems \citep{Me09mn} predicts a radius of 0.41\Rsun\ -- but is in excellent agreement with the semi-empirical sequence for CV secondary stars constructed by \citet{Knigge06mn}. This is expected because the mass transfer timescale becomes similar to the thermal timescale for the secondary components of CVs above the period gap, allowing continued mass transfer to drive the star out of thermal equilibrium.

\subsection{Distance and white dwarf temperature}

\begin{figure} \centering
\includegraphics[angle=-90,width=\columnwidth]{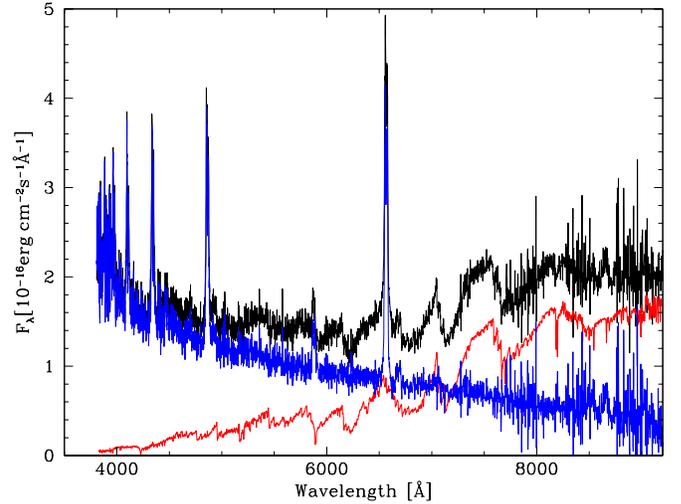}
\caption{\label{f-sp2} Black line: the SDSS spectrum of SDSS\,J1006.
Red line: an M3.2 template spectrum scaled to match the strengths of the
spectral features of the companion star in SDSS\,J1006. Blue line: the
residual spectrum obtained after subtracting the M-dwarf template from
the spectrum of SDSS\,J1006.} \end{figure}

\begin{figure} \centering
\includegraphics[angle=-90,width=\columnwidth]{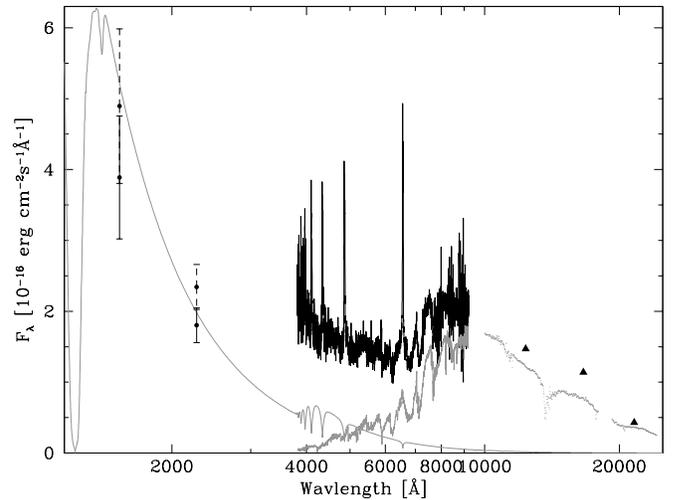}
\caption{\label{f-sed} Spectral energy distribution of SDSS\,J1006,
showing the observed (solid error bars) and dereddened ($\EBV = 0.03$;
dashed error bars) GALEX UV fluxes, the optical SDSS spectrum, and the
2MASS near-IR fluxes (black triangles). Shown in gray are the M3.2 SDSS
template from Fig.\,\ref{f-sp2}, near-IR fluxes for an M3.2 star
interpolated from Sandy Leggett's archive of M-dwarf spectra, and a
$\logg = 7.93$, $T_\mathrm{WD} = 16\,500$\,K model spectrum scaled for
$R_\mathrm{WD} = 0.016$\Rsun\ and $d = 676$\,pc.} \end{figure}

The secondary star dominates the red end of the optical spectrum of SDSS\,J1006, exhibiting the strong TiO bandheads that are characteristic of mid-to-late M dwarfs. We have obtained the star's spectral type using the M dwarf template library that \citet{Rebassa+07mn} assembled from SDSS spectroscopy, interpolated onto a finer grid spanning types M0 to M9 in steps of 0.2 subtypes. These templates were scaled and subtracted from the SDSS spectrum of SDSS\,J1006 until the smoothest residual spectrum was obtained, resulting in a spectral type of M3.2$\pm$0.2 (Fig.\,\ref{f-sp2}). We find a good agreement with the relationship between spectral type and mass presented by \citet{Rebassa+07mn}. From the best-fit template, we calculated $f_\mathrm{TiO}$, which is the flux difference between the bands $7450$--$7550$\,\AA\ and $7140$--$7190$\,\AA, as defined by \citet{Beuermann06aa}. Using the polynomial expressions of \citet{Beuermann06aa}, we obtained the $F_\mathrm{TiO}$ surface fluxes for secondary stars with spectral types in the range M3.0 to M3.4. Taking $R_2=0.466$\Rsun, we find a distance of
\[  d = \sqrt{\frac{R_2^2F_\mathrm{TiO}}{f_\mathrm{TiO}}} = 676 \pm 40 \,\mathrm{pc}  \]
where the uncertainty in $d$ is dominated by that in $R_2$.

SDSS\,J1006 has also been detected in the ultraviolet (UV) all-sky survey carried out by GALEX \citep{Martin+05apj}, from which we can estimate the WD's effective temperature, $T_{\rm WD}$. The GALEX observations were obtained during the phase interval 0.117--0.124, which is outside the WD eclipse to high confidence. We adopt $\logg = 7.93$ for the WD (Table\,\ref{tab:final}), and $d = 676$\,pc. \reff{Under the assumption that the UV flux is entirely due to the unobscured photospheric emission of the WD,} the only free parameter to reproduce the observed GALEX fluxes is then $T_{\rm WD}$. The observed far-UV flux implies $T_{\rm WD} \approx 16\,000$\,K, or $T_{\rm WD} \approx 17\,000$\,K if a maximum reddening of $\EBV = 0.031$ \citep{Schlegel++98apj} is assumed (Fig.\,\ref{f-sed}), with an uncertainty of $\pm$1500\,K. \reff{We therefore adopt $T_{\rm WD} = 16\,500 \pm 2000$\,K. This value is only an estimate, because it is possible that the WD is partially veiled by the accretion disc, and that the disc, bright spot and boundary layer contribute to the observed UV fluxes. A more reliable $T_{\rm WD}$ measurement could be obtained from UV spectroscopy.}

This $T_{\rm WD}$ is unusually low for a dwarf nova with $\Porb \sim 4$\,hr \citep[e.g.][]{UrbanSion06apj, TownsleyGansicke09apj} -- by comparison the well-studied dwarf nova U\,Gem has overall properties which are very similar to SDSS\,J1006 but harbours a WD with $T_{\rm WD} \approx 30\,000$\,K \citep{Sion+98apj,Long++06apj}. Given that SDSS\,J1006 is a high-inclination system, it might be possible that the WD is partially veiled by extended structures above the accretion disc, similar to those seen in OY\,Car \citep{Horne+94apj}. If veiling is not the cause of the low observed UV flux (no veiling is observed in U\,Gem), the low $T_{\rm WD}$ implies a secular mean accretion rate of a few $10^{-10}$\Msun\,yr$^{-1}$ \citep{TownsleyGansicke09apj}, which is a factor of about three lower than in U\,Gem.

\subsection{Outburst characteristics}

% $g = 18.34$
% $gspec = 18.63$
% GALEX phase 0.127 $\pm$ 0.048
% J = 15.833 +/- 0.066
% H = 14.976 +/- 0.072
% K = 14.998 +/- 0.110
% JD of 2MASS measurements: 2450838.8896, so phase 0.15 $\pm$ 0.04

\begin{figure} \centering
\includegraphics[angle=0,width=\columnwidth]{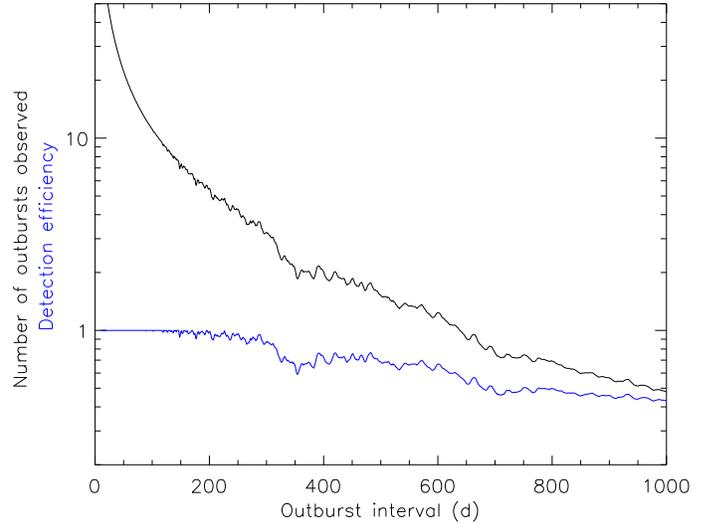}
\caption{\label{fig:outburst} Outburst detection efficiency for the Catalina
Sky Survey observations of SDSS\,J1006 (blue lower line) and the number of
\reff{outbursts} which would have been detected for SDSS\,J1006 (black upper
line), as a function of the outburst period.} \end{figure}

SDSS\,J1006 has been observed by the Catalina Sky Survey \citep{Drake+09apj}, who obtained 209 unfiltered magnitude measurements between 2005 April and 2008 April\footnote{See {\tt http://nesssi.cacr.caltech.edu/catalina/20050301/ \\ SDSSCV.html\#table77}}. Most of these observations found the object in the magnitude interval 17--18, but two dwarf nova outbursts have also been observed (at JDs $2453678$ and $2454259$). \reff{The durations of these outbursts are not known, but are constrained to be more than one day in the first case.}

We investigated the dwarf nova outburst frequency of SDSS\,J1006, using Monte Carlo simulations and the times of the Catalina observations. Assuming that each outburst is observable for a 10\,d period \citep{Ak++02aa}, we obtained a detection efficiency of 39\% over the full time span of the Catalina observations. If we further assume that the outbursts occur periodically, we can obtain the detection efficiency (and thus the probable number of outbursts observed) as a function of outburst frequency. The results of this calculation are shown in Fig.\,\ref{fig:outburst} and favour an outburst interval in the region of 400\,d. This is a very long interval for a 4--5\,hr period CV: \citet{Ak++02aa} find a mean outburst interval of 62.0\,d for U\,Gem-type systems.

Based on its photometric and spectroscopic properties, SDSS\,J1006 can be classified as a dwarf nova of U\,Gem type, for which two dwarf nova outbursts have been detected and the outburst interval is long. Continued observations of this object would be very useful in refining its outburst frequency.

%%%%%%%%%%%%%%%%%%%%%%%%%%%%%%%%%%%%%%%%%%%%%%%%%%%%%%%%%%%%%%%%%%%%%%%%%%%%%%%%%%%%%%%%%%%%%%%%%%%%%%%%%%%%%%

\section{Summary and discussion}

\begin{table*} \centering \caption{\label{tab:catalogue} Eclipsing cataclysmic variables for
which masses and radii of one or both components has been measured accurately and precisely.
Those secondary radii without errorbars are not available from the original reference so have
instead been calculated assuming Roche geometry.
\newline {\bf References:} (1) \citet{Ribeiro+07aa}; (2) Copperwheat et al.\ in preparation;
(3) \citet{Gansicke+00aa}; (4) \citet{Thorstensen00pasp}; (5) \citet{Feline+05mn};
(6) \citet{Marsh+90apj}; (7) \citet{LongGilliland99apj}; (8) \citet{Naylor++05mn};
(9) \citet{Echevarria++07aj}; (10) \citet{Horne++93apj}; (11) \citet{Wood+05apj};
(12) \citet{Fiedler++97aa}; (13) \citet{Baptista++00mn}; (14) \citet{Baptista++01mn};
(15) \citet{Thoroughgood+05mn}; (16) \citet{Thoroughgood+04mn}. }
\begin{tabular}{l c r@{\,$\pm$\,}l r@{\,$\pm$\,}l r@{\,$\pm$\,}l r@{\,$\pm$\,}l r@{\,$\pm$\,}l l} \hline
Name&Orbital    &\mc{Mass ratio}&\mc{White dwarf} &\mc{White dwarf}   &\mc{Secondary}   &\mc{Secondary}     &References\\
    &period (d) &\mc{ }         &\mc{mass (\Msun)}&\mc{radius (\Rsun)}&\mc{mass (\Msun)}&\mc{radius (\Rsun)}&          \\
\hline
IP\,Peg           & 0.158206 & 0.48 & 0.01 & 1.16 & 0.02 &0.0064 & 0.0004& 0.55 & 0.02 & 0.47  & 0.01 & 1, 2      \\
GY\,Cnc           & 0.175442 & 0.387&0.031 & 0.99 & 0.12 &    \mc{ }     & 0.38 & 0.06 &   \mc{0.44}  & 3, 4, 5   \\
U\,Gem            & 0.176906 & 0.362&0.010 & 1.14 & 0.07 &0.0067 & 0.0008& 0.41 & 0.02 & 0.43  & 0.06 & 6, 7, 8, 9\\
SDSS\,J1006$+$2337& 0.185913 & 0.51 & 0.08 & 0.78 & 0.12 &0.016  & 0.006 & 0.40 & 0.10 & 0.466 & 0.036& This work \\
DQ\,Her           & 0.193621 & 0.62 & 0.05 & 0.60 & 0.07 &    \mc{ }     & 0.40 & 0.05 & 0.49  & 0.02 & 10, 11    \\
EX\,Dra           & 0.209937 & 0.75 & 0.01 & 0.75 & 0.02 &0.013  & 0.001 & 0.56 & 0.02 & 0.57  & 0.02 & 12, 13, 14\\
V347\,Pup         & 0.231936 & 0.83 & 0.05 & 0.63 & 0.04 &    \mc{ }     & 0.52 & 0.06 & 0.60  & 0.02 & 15        \\
EM\,Cyg           & 0.290909 & 0.88 & 0.05 & 1.13 & 0.08 &    \mc{ }     & 0.99 & 0.12 & 0.87  & 0.07 & 15        \\
AC\,Cnc           & 0.300477 & 1.02 & 0.04 & 0.76 & 0.03 &    \mc{ }     & 0.77 & 0.05 &   \mc{0.83}  & 16        \\
V363\,Aur         & 0.321242 & 1.17 & 0.07 & 0.90 & 0.06 &    \mc{ }     & 1.06 & 0.11 &   \mc{0.90}  & 16        \\
\hline \end{tabular} \end{table*}

From the observations presented in this work we have discovered that SDSS\,J1006 is an eclipsing CV, measured the orbital period, and calculated the masses and radii of its component stars. This was achieved through a parameteric model of its eclipses, \reff{combined with} a spectroscopic velocity amplitude for the secondary star corrected for the effects of irradiation. Doppler maps of the infrared calcium triplet reveal emission from the secondary star and are in agreement with this $K_{\rm MD}$. From the spectral characteristics of the system we have also found a WD effective temperature of $T_{\rm WD} = 16\,500 \pm 2000$\,K, a secondary star spectral type of M3.2$\pm$0.2, and a distance of $d = 676 \pm 40$\,pc. A dwarf nova outburst interval of roughly 400\,d agrees with the available photometric observations of SDSS\,J1006.

We measured radial velocities for the WD from the H$\alpha$ and H$\beta$ emission lines, finding a velocity amplitude $K_1 = 127.1 \pm 4.2$\kms. Despite using the diagnostic diagram approach, our RVs still have a phase offset of 0.04 from the eclipse ephemeris. We therefore could not assume that they represent the motion of the WD, so did not use them in obtaining the physical properties of SDSS\,J1006. Notwithstanding this, the $K_1$ we measured turns out to be in excellent agreement with the {\em expected} WD velocity amplitude of 131.5\kms.

The mass of the WD is $0.78 \pm 0.12$\Msun, and its radius is consistent with theoretical expectations. The secondary star has a mass and radius of $0.40 \pm 0.10$\Msun\ and $0.466 \pm 0.036$\Rsun, respectively, which is in excellent agreement with the semi-empirical sequence for CV secondary stars constructed by \citet{Knigge06mn}. The uncertainties in the system parameters are dominated by the moderate quality of the light curve, and an improved photometric study of this object is warranted.

In Table\,\ref{tab:catalogue} we have assembled a list of the component masses and radii of eclipsing CVs with long orbital periods (greater than 3.0\,hr). We discount systems with uncertain properties or whose analysis rests on emission-line RVs (not always reliable) or mass--radius relations for the secondary star (to avoid circular arguments). The list is worryingly short: only 10 systems (including SDSS\,J1006) satisfy our criteria, of which one is magnetic (DQ\,Her). The weighted mean and standard deviation of the WD masses is $0.78 \pm 0.19$\Msun. The masses and radii of the secondary stars display a clear negative correlation with orbital period, as expected by our current understanding of the evolution of CVs. The WD masses display no correlation with orbital period or with the secondary star masses, in agreement with studies which show that WDs in CVs do not undergo large overall changes in mass \citep{PrialnikKovetz95apj,Knigge06mn}.

%%%%%%%%%%%%%%%%%%%%%%%%%%%%%%%%%%%%%%%%%%%%%%%%%%%%%%%%%%%%%%%%%%%%%%%%%%%%%%%%%%%%%%%%%%%%%%%%%%%%%%%%%%%%%%

\begin{acknowledgements}

The reduced observational data presented in this work will be made available at the CDS ({\tt http://cdsweb.u-strasbg.fr/}) and at {\tt http://www.astro.keele.ac.uk/$\sim$jkt/}. \ We are grateful to Andrew Drake for providing the Catalina Sky Survey observations of SDSS\,J1006, and to the anonymous referee for a positive report. JS, TRM, BTG and CMC acknowledge financial support from STFC in the form of grant number ST/F002599/1. ARM acknowledges financial support from ESO, and Gemini/Conicyt in the form of grant number 32080023. \ Based on observations made with the William Herschel Telescope, operated by the Isaac Newton Group, and the Nordic Optical Telescope, operated jointly by Denmark, Finland, Iceland, Norway, and Sweden, both on the island of La Palma in the Spanish Observatorio del Roque de los Muchachos of the Instituto de Astrof\'{\i}sica de Canarias. \reff{Based on observations collected at the Centro Astron\'omico Hispano Alem\'an (CAHA) at Calar Alto, Spain, operated jointly by the Max-Planck Institut f\"ur Astronomie and the Instituto de Astrof\'{\i}sica de Andalucía (CSIC).} The following internet-based resources were used in research for this paper: the ESO Digitized Sky Survey; the NASA Astrophysics Data System; the SIMBAD database operated at CDS, Strasbourg, France; and the ar$\chi$iv scientific paper preprint service operated by Cornell University.

\end{acknowledgements}

%%%%%%%%%%%%%%%%%%%%%%%%%%%%%%%%%%%%%%%%%%%%%%%%%%%%%%%%%%%%%%%%%%%%%%%%%%%%%%%%%%%%%%%%%%%%%%%%%%%%%%%%%%%%%%

% \bibliographystyle{aa}
% \bibliography{jkt}

%%%%%%%%%%%%%%%%%%%%%%%%%%%%%%%%%%%%%%%%%%%%%%%%%%%%%%%%%%%%%%%%%%%%%%%%%%%%%%%%%%%%%%%%%%%%%%%%%%%%%%%%%%%%%%
\end{document}